\documentclass[a4paper]{article}
\usepackage{amssymb,graphicx,url,amsmath,subfigure
}
\usepackage{times}
\usepackage{tikz}
\usepackage{soul}
\usepackage{color}
\usepackage{xcolor}
\usepackage{algorithm}
\usepackage{algorithmic}
\usepackage{enumerate}
\usepackage{a4wide}
\usepackage{comment}
\usetikzlibrary{shapes.geometric,positioning}

\newcommand{\one}{{\bf 1}}

\newcommand{\cA}{{\cal A}}

\newcommand{\R}{{\mathbb R}}
\newcommand{\N}{{\mathbb N}}
\newcommand{\C}{{\mathbb C}}
\newcommand{\Z}{{\mathbb Z}}

\newtheorem{Theorem}{Theorem}

\definecolor{MyDarkGreen}{rgb}{0.17,0.46,0.25} 
\definecolor{MyDarkRed}{rgb}{0.88,0.22,0.21} 
\definecolor{MyDarkBlue}{rgb}{0.11,0.11,0.70}

\definecolor{lightgray}{gray}{0.85}
\sethlcolor{Dandelion}

\usetikzlibrary{arrows,shapes,plotmarks,positioning}
\usetikzlibrary{decorations.markings}

\tikzset{>=stealth'} 
\tikzstyle{graphnode} = 
   [circle,draw=black,minimum size=22pt,text centered,text
     width=22pt,inner sep=0pt] 
\tikzstyle{var}   =[graphnode,fill=white]
\tikzstyle{vardashed}   =[graphnode,draw=gray,fill=white]
\tikzstyle{obs}   =[graphnode,fill=black,text=white]
\tikzstyle{obsgrey}   =[graphnode,draw=white,fill=lightgray,text=black]
\tikzstyle{par}    =[graphnode,draw=white,fill=red,text=black] 
 \tikzstyle{crucial} =[graphnode,draw=white,fill=yellow,text=black] 
\tikzstyle{fac}   =[rectangle,draw=black,fill=black!25,minimum size=5pt]
\tikzstyle{facprior} =[rectangle,draw=black,fill=black,text=white,minimum size=5pt]
\tikzstyle{edge}  =[draw=white,double=black,very thick,-]
\tikzstyle{blueedge}  =[draw=white,double=blue,very thick,-]
\tikzstyle{rededge}  =[draw=white,double=red,very thick,-]
\tikzstyle{prior} =[rectangle, draw=black, fill=black, minimum size=
5pt, inner sep=0pt]
\tikzstyle{dirprior} = [circle, draw=black, fill=black, minimum
size=5pt, inner sep=0pt]

\date{April 15, 2018}

\title{Simple negative result for physically universal controllers with macroscopic interface}

\author{Dominik Janzing\\
{\small dominik.janzing@tuebingen.mpg.de} \\
       {\small Max Planck Institute for Intelligent Systems}\\
       {\small Max-Planck-Ring 4}\\
       {\small 72076 T\"ubingen, Germany}
}
\begin{document}
\maketitle

\abstract{To study potential limitations of 
controllability of physical systems I have
earlier proposed {\it physically universal} cellular automata and Hamiltonians. These are translation invariant interactions for which any control operation on a finite target region can be implemented by the autonomous time evolution 
if the complement of the target region is `programmed' to an appropriate initial state.  
This provides a model of control
where the cut between a system and its controller can be consistently shifted,  in analogy to the Heisenberg cut 
defining the boundary between a quantum system and its measurement device. 
However, in the known physically universal CAs 
the implementation of microscopic 
transformations requires to write the `program' into {\it microscopic} degrees of freedom, while human actions take place on the {\it macroscopic} level. 
I therefore ask whether there exist physically universal interactions for which any desired operation on a target region can be performed by only controlling the macroscopic state of its surrounding.
A very simple argument shows that this is impossible
with respect to the notion of `macroscopic' proposed here:
control devices whose position is only
specified up to `macroscopic precision'
cannot operate at a precise location in space. 
This suggests that reasonable notions of
`universal controllability' need to be tailored to the manipulation of {\it relative} coordinates, but it is not obvious how to do this. The statement that any microscopic transformation can be implemented in principle, whenever it is true in any sense, it does not seem to be true in its most obvious sense. 
}

\section{Motivation}
During the past decades significant progress has been made regarding the ability of controlling elementary quantum systems \cite{NC}. Experiments that are meanwhile feasible include 
manipulating single trapped ions \cite{Schindler2013} and also preparing coherent superpositions, interference experiments with
large atoms \cite{Arndt2005}, or controlling interacting spins in nuclear magnetic resonance \cite{Boykin}, 
just to mention a few examples. These success stories raise a variety of fundamental questions: 
  
\paragraph{Is any unitary operation feasible?}
In standard quantum mechanics textbooks, skeptical remarks about whether any unitary on arbitrary system Hilbert spaces can be implemented in reality, are rare.
Indeed, in the context of quantum computing \cite{NC},
researchers have studied a broad variety of quantum systems for which any unitary can be approximated by 
concatenating elementary transformations \cite{DiV} whose implementation has been demonstrated. 
At first glance, it seems that the question of feasibility
of an arbitrary unitary thus reduces to
complexity theoretic issues
 -- which would amount to {\it fundamental} limitations `only' after accounting for finiteness
of time resources, e.g.,  
due to the potential finite life time of the universe. The feasibility of complex transformations, however, could even be restricted if infinite implementation time is allowed, namely if every elementary unitary can only be implemented up to
some tiny error accuracy. Contrary to a common belief,
tiny errors do matter for this question {\it despite} the existence of error correcting codes \cite{Calderbank}. This is because
error correcting codes only ensure the ability
to implement any unitary on the {\it logical} space, while
a large part of the {\it physical} state space cannot be reached. 
This will be briefly explained in the following paragraph. 

\paragraph{Emergence of a classical world}
The contrast between quantum theory and our every day life being governed by classical laws, has been subject of debates 
since the early days of quantum mechanics, see e.g. \cite{Omnes,Guilini96} and references therein. 
After recent results on the foundations of
quantum theory \cite{Spekkens2016} have provided
a more profound understanding about which phenomena 
are genuinely quantum and which ones can also be understood classically \cite{Spekkens2016b}, 
it may be recommended to reconsider the emergence
of a classical world with more conceptual clarity. 
Assuming that any formally possible measurement and unitary can be implemented in principle implies the measurability of observables that are incompatible with observables that are believed to be {\it classical}, which challenges the existence of a classical world and the irreversibility of
the measurement process. 

To explain possible limitations for the implementation of
arbitrary macroscopic superpositions, we consider 
a class of observables that are good candidates for being particularly `classical'.
To this end, consider a system that consists of $n$ qubits, just to have a simple example. For any self-adjoint operator $a$ acting on $\C^2$, let
\[
a_j:= \underbrace{\one \otimes \cdots \otimes \one}_{j-1} \otimes a \otimes \one \otimes \cdots \otimes \one 
\]
be its copy acting on qubit $j$.
Then we define
the mean field observable
\begin{equation}\label{eq:meanfield}
\overline{a}:=\frac{1}{n} \sum_{j=1} a_j.
\end{equation}
For instance, $\overline{\sigma_z}$ in a spin chain, would be the average spin in $z$-direction, an observable that is for 
large $n$ directly accessible to human experience via its 
magnetization. The observables 
$\overline{\sigma_x},\overline{\sigma_y},\overline{\sigma_z}$ can be approximately measured simulatenously since they almost commute  
 \cite{Poulin}, suggesting to consider 
the vector $(\overline{\sigma_x},\overline{\sigma_y},\overline{\sigma_z})$ as the classical magnetization. 
 Yet, this view is problematic if one assumes that
the `Schr\"odinger cat state' 
\begin{equation}\label{eq:cat}
\psi:=\frac{1}{\sqrt{2}}\left(|0\cdots 0\rangle  +
|1\cdots 1\rangle \right)
\end{equation}
can be prepared. Here, we have 
assumed that we are able to measure the observables 
that distinguish \eqref{eq:cat} from the mixture of
$|0\cdots 0\rangle$ and $|1\cdots 1\rangle$
(otherwise it would be unclear how to define 
the difference between coherent superposition and mixture anyway unless one commits to an ontic interpretation of the wave function). 
To discuss possible limitations for preparing states like \eqref{eq:cat}, note that \cite{JB00} has shown the following impossibility result:
whenever one starts in a product state and the preparation procedure relies on one- and two-qubit gates which 
are imprecise in the sense that they are slightly depolarizing
 (i.e., each gate 
outputs the maximally mixed state with some tiny probability $\epsilon$), states like \eqref{eq:cat}
cannot be approximated in principle. 
If one believes that $\epsilon$ can get arbitrarily close to zero when technology advances, the results in \cite{JB00} do not imply any {\it fundamental} restrictions. However, theoretical results suggest 
fundamental lower bounds on the error rates
that are based on the quantumness of the controllers, for instance, due to the finiteness of the clocking devices \cite{Gambini2004}.
Until we understand how the interface between our classical actions and microscopic degrees of freedom really works, we cannot decide the potential limitations to controllability. 
Although these remarks are speculative, they  nevertheless motivate the construction of models 
of control within which potential limitations can be derived. 


\paragraph{Emergence of accuracy}
Regardless of the precision that is achieved by 
modern control technology, any human action on the microscopic world is finally performed by 
macroscopic actions on devices that are large enough to be grasped with our hands. Remarkably, the inaccuracy that is inherent to any motion of our hands, is not necessarily inherited by the control operation (for instance, if
the tip of an 	scanning tunneling microscope is moved on the scale of single atoms).
In other words, we are able to {\it precisely} control
tiny objects by {\it imprecisely} moving macroscopic objects, raising the question how this accuracy `emerges'.\footnote{See also the debate about limitations of future nanotechnology and whether one will ever produce nanorobots that 
reproduce themselves \cite{Smalley2001}. The problem  of controlling tiny things like atoms with `fat and sticky' fingers is already mentioned there.} 

\begin{figure}
\includegraphics[width=\textwidth]{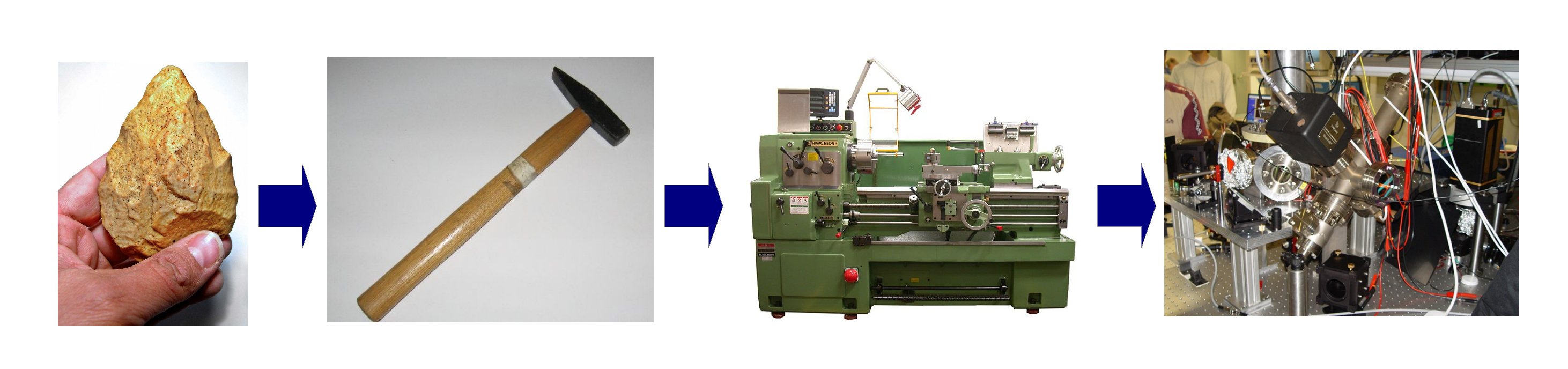}
\caption{\label{fig:evolution} Sketch of the evolution of precision in technological evolution, from the left to the right: Hand\-axe, Hammer, turning machine, ion trap. Humans used primitive tools to produce tools with increasing precision. Obviously, the tools do not necessarily inherit the inaccuracy from the tools that have been used for producing them.} 
\end{figure}

There is also another kind of `emergence of accuracy' in time
that is given by technological evolution, see Figure~\ref{fig:evolution}\footnote{Images taken from Wikipedia, authors (from left to right): (1) Theroadislong, (2) unspecified (3) Glenn McKechnie, (4) Mnolf. Resdistribution only under licenses described there (keywords:
`handaxe', `Hammer', `Drehmaschine', `ion  trap').}.
While handaxes from the stone age would not have been
appropriate to precisely control tiny quantum systems, those stone age tools have obviously been powerful enough to produce more precise tools, which, in turn, were able to produce even more precise ones. It sounds paradox that a tool can be more precise than the tool it has been manufactured by, but obviously this is possible. 
Since it seems that such a process involves a large number of steps in between (one couldn't have produced the ion trap directly by an handaxe),  one can speculate that operating on microscopic degrees of freedoms by acting on macroscopic ones involves some logical depth
in the computational model provided below.\footnote{The idea that intuitive notions
of complexity of physics are associated with logical depth can already be found in \cite{BennettDepth}.}

These remarks are, again, quite vague and only supposed
to motivate models of controllers for microscopic systems that explain `emergence of accuracy' in the sense that
microscopic degrees of freedom can be accessed by operating on macroscopic ones.

\paragraph{Structure of the paper}
Subsection~\ref{subsec:PhysUniv} sketches the
notion of physically universal CAs and Hamiltonians
introduced in \cite{PhysUniversal} and explains in what sense they can be seen as a toy model for a world
in which every physical degree of freedom is indirectly accessible by operating on the remaining degrees of freedom. 
Subsection~\ref{subsec:macro} shows that it is impossible to control any microscopic degree of freedom by only
specifying the `macroscopic' properties
 of the regions surrounding the target region. 
The fact that it is in fact feasible to control
microscopic degrees of freedoms by human actions that operate on the macroscopic scale suggests
that universal controllability
can only hold in a weaker sense.


\section{Universal controllers with macroscopic interface}

\subsection{Physical universality\label{subsec:PhysUniv}}

I first introduce the concept of physical universality as introduced in \cite{PhysUniversal} and explain what qualifies physically universal CAs and Hamiltonians for studying how to
control physical degree of freedoms. 
This point can be made without being overly formal. 

\paragraph{Classical CAs}
A classical cellular automaton in dimension $d$
consists of cells corresponding to points in the lattice $\Lambda:=\Z^d$, each cell can attain values in some finite alphabet $\Sigma$. The discrete dynamics $\alpha$ is given by some local translational invariant update rule that is repeated in every time step. Before we explain physical universality, we restrict the attention to
{\it reversible} CAs, since this condition is necessary for physical universality as observed by Schaeffer \cite{Schaeffer2015}. The simplest
way to ensure reversibility is given by so-called Margolus neighborhoods \cite{Marg}. This is an update scheme consisting of two 
update rules for odd and even time steps, respectively.  In two dimensions, for instance, 
one update rule is a transformation acting on each square 
\[
S^{i,j}_{\rm even}:=((2i,2j),(2i+1,2j),(2i,2j+1),(2i+1,2j+1)) \quad i,j \in \Z,
\]
 while the other one acts on the square
\[
S^{i,j}_{\rm odd}:=((2i-1,2j-1),(2i,2j-1),(2i-1,2j),(2i,2j)) \quad i,j \in \Z.
\] 
In the even and odd time step, the update rules permute the configurations
in $\Sigma^{S^{i,j}_{\rm even}}$ and $\Sigma^{S^{i,j}_{\rm odd}}$, respectively.  
This resulting update rule $\alpha$ then is a bijection of configurations
in $\Sigma^\Lambda$. The overall update rule defines
the autonomous time evolution of the system.

We now define what it means that this time evolution 
`implements' some desired transformation.
For any finite region $R\subset \Lambda$ let $\beta:\Sigma^R\to \Sigma^R$ be a map on the possible configurations of $R$. A configuration $c\in \Sigma^{\Lambda \setminus R}$ of the remaining part of the lattice is said to {\it implement} $f$ after time $t\in \N$
if 
\[
\alpha^t\left((c,r)\right)|_R = \beta (r),
\]
for every configuration $r\in \Sigma^R$, with $|_R$ denoting the restriction of the output configuration to $R$. Here $(c,r)$ denotes the 
configuration of the entire lattice defined
by the restrictions $c$ and $r$ to $\Lambda \setminus R$ and $R$, respectively.  
The CA is said to be {\it physically universal} if there is
a configuration of $\Lambda \setminus R$ implementing $f$ for any function $f$ 
and any finite region $R$.
Meanwhile, a physically universal classical CA for $d=2$ has been constructed by Schaeffer \cite{Schaeffer2015}, and for $d=1$ by
Salo \& T\"orm\"a \cite{Salo2015}.

\paragraph{Quantum CAs}
The natural analog of the classical CA above
is a lattice whose cells contain the Hilbert space
$\C^{|\Sigma|}$, with basis vectors labelled by the alphabet $\Sigma$.
The observable algebra of a single cell is given by $M(\C^{|\Sigma|})$. To avoid
issues of infinite tensor products of Hilbert spaces, mathematical physics uses an algebraic framework that I briefly mention for sake of completeness, although its details are not important for the problems discussed here.
Readers who are not familiar with this framework
may also skip these details and think of a state of the infinite lattice simply as a family of 
density matrices for finite regions whose restrictions to common subsets of cells are consistent. 

For any finite region $R$, the observable algebra
$\cA_R$ is given by the tensor product of $|R|$ copies of 
 $M(\C^{|\Sigma|})$, with the canonical embedding that considers $\cA_R$ as subalgebra of $\cA_{R'}$ for $R\subset R'$. 
 Then the algebra $\cA_{\Lambda}$ of the entire CA is given by the $C^*$-inductive limit of these algebras $\cA_R$ \cite{WernerCA}, also called the quasi-local algebra \cite{BR1}. 
It is given by the completion of the union
$\cup_R \cA_R$ over all finite regions $R$
with respect to the operator norm. 
A state $\rho$ is no longer be given by a vector or a density operator -- since we are not talking about the Hilbert space of the infinite system, but only about the algebra of observables. Instead, it is given by a linear positive functional $\rho:\cA_\Lambda \to \R$ of norm $1$   
\cite{Mu90}. With any (not necessarily) finite subset $S\subset \Lambda$ we can then associate a $C^*$-subalgebra $\cA_S$ of $\cA_\Lambda$ generated 
  by subalgebras of finite regions in $S$.
  
Reversible update rules are, again, most easily constructed 
using Margolus neighborhoods, where independent unitaries act on each of the above quadruples of cells. 
They define $C^*$-automorphisms  \cite{Mu90,BR1} for each
$\cA_{S^{i,j}_{\rm even}}$ and $\cA_{S^{i,j}_{\rm odd}}$, respectively, which results in an $C^*$-algebra automorphism of $\cA_{\Lambda}$. Its dual is a map on the states of $\cA_\Lambda$
that we also
call $\alpha$ (like the update rule in the classical CA). This overloading of notation will be convenient if we later talk about the behaviour of $\alpha$ in cases where we do not need to distinguish between the quantum and the classical case.
The state $\rho$ of $\cA_{\Lambda \setminus R}$ 
is said to {\it implement} the unitary $u\in \cA_R$ 
up to the accuracy $\epsilon$
whenever
\[
\|\alpha(\rho \otimes \gamma )|_R - u\gamma u^\dagger\|_1 \leq \epsilon,
\]
for any state $\gamma$ on $\cA_R$. We call a state $\rho$ a `basis state' whenever its restriction to any finite region $R$
is given by
\[
\rho|_R (b) =\langle \psi, b\, \psi\rangle,
\]
with $\psi$ being a tensor product of single cell basis states:
\[
|\psi\rangle :=\otimes_{\lambda \in R} |s_\lambda \rangle, 
\]
with $s_\lambda \in \Sigma$. 
A quantum CA is called 
{\it physically universal} if for any finite target region
$T\subset \Lambda$ of cells, and any unitary
$u\in \cA_T$,
there is a basis state $\rho$ of $\cA_{\Lambda\setminus T}$ 
 and a time $t\in \N$ such that the $t$-fold application of the update rule $\alpha$ implements $u$ for any desired accuracy $\epsilon$. A physically universal
quantum CA for $d=2$ has been constructed by
Schaeffer \cite{Schaeffer2015quantum}.

\paragraph{Replacing the CA with a  Hamiltonians}
 To get closer to physics, one may want to replace the discrete time evolution by a
fixed translational invariant Hamiltonian
thus resembling the usual notion of spin chains 
\cite{Vollbrecht}. Then the time evolution is a group $(\alpha_t)_{t\in \R}$ of $C^*$-automorphisms obtained by the limit of automorphisms induced by Hamiltonians $H_R$ that act on finite regions only, a construction that is standard for describing the dynamics of infinite spin chains in mathematical physics \cite{BR2}. 

Below we will only use the discrete setting for sake of its simplicity. After all, discrete update rules close to the identity can
approximate the Hamiltonian evolution anyway, which shows that the discrete idealization is not too far from physics.

\paragraph{Why studying physical universality?}
To explain why known constructions of  CAs \cite{Shepherd} are not sufficient for our purpose, note
that they admit universal control  on the {\it data} cells only, with the program being written into 
distinct {\it program} cells. This distinction is not allowed if one also wants to understand how to manipulate the data cells. 
In other words, a model of controlling microphysics should not only be {\it computationally} universal in the sense of being able to act universality one some {\it logical} subspace. Instead, it is supposed to act universally on the entire {\it physical} space.\footnote{For possible
thermodynamic consequences of
this strong notion of universality see \cite{cycles}.} The following thoughts experiment may justify physical universality even further.
Assume we believe that any physical degree of freedom is controllable, that is, there are human actions on
some control devices that ensure that the desired operation is performed on the target system. It should then be possible 
to build a technical device that replaces the human and implements the actions on the control device automatically. This setup would involve some robot-like actors replacing the motion of human hands, plus some computing device replacing the human cognitive process 
required to properly act on the control device (including a correct timing of those actions).
We now consider an `overall' device consisting of all these devices together. The dynamics of the overall
device is given by some fixed interaction of the physical particles it consists of. 
If our CA is supposed to be a toy model of the world 
--in which presence and absence of matter is described by 
appropriate states of an a priori homogeneous space-- both building the hardware of the controlling device, and programming its program register, just amounts to writing a program into the CA. These arguments suggest to use physically universal interactions as models, given that we belief that 
every formally possible control operation
can be performed by humans.\footnote{For other notions of universality, e.g., those that refer to the ability of a machine to reproduce itself, see also von Neumann's universal constructor \cite{vonNeumann1966}, cp. also constructor theory by Deutsch et al. \cite{constructor}.} 




\subsection{The problem of specifying only macroscopic properties \label{subsec:macro}}
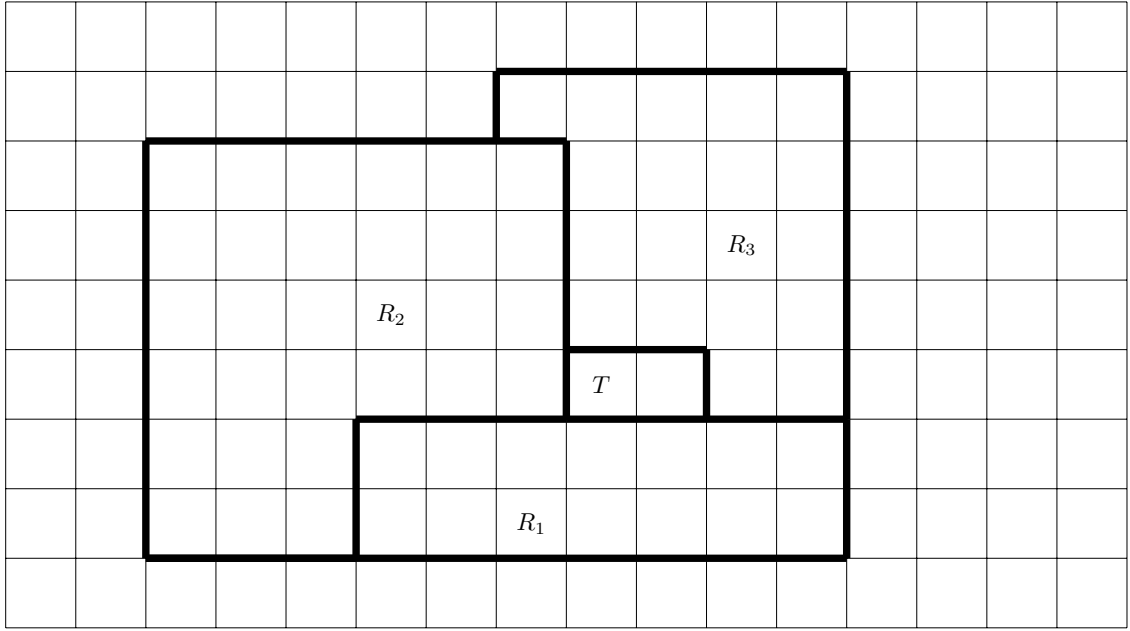
\begin{figure}
\centerline{
\resizebox{\textwidth}{!}{
\begin{tikzpicture}
\draw[step=1cm,black,very thin] (-8,-3) grid (8,6);
\coordinate (a) at (0,0);
\coordinate (b) at (2,0);
\coordinate (c) at (2,1);
\coordinate (d) at (0,1);
\draw [line width=3pt] (a) -- (b);
\draw [line width=3pt] (b) -- (c);
\draw [line width=3pt] (c) -- (d);
\draw [line width=3pt] (d) -- (a);
\coordinate (e) at (-3,0);
\draw [line width=3pt] (a) -- (e);
\coordinate (f) at (-3,-2);
\draw [line width=3pt] (e) -- (f);
\coordinate (g) at (0,4);
\draw [line width=3pt] (d) -- (g); 
\coordinate (h) at (-6,4);
\draw [line width=3pt] (g) -- (h);
\coordinate (i) at (-6,-2);
\draw [line width=3pt] (i) -- (h);
\draw [line width=3pt] (i) -- (f);
\coordinate (j) at (4,-2);
\draw [line width=3pt] (j) -- (i);
\coordinate (k) at (4,0);
\draw [line width=3pt] (j) -- (k);
\draw [line width=3pt] (k) -- (b);
\coordinate (l) at (4,5);
\draw [line width=3pt] (l) -- (k);
\coordinate (m) at (-1,5);
\draw [line width=3pt] (l) -- (m);
\coordinate (n) at (-1,4);
\draw [line width=3pt] (n) -- (m);
\node at (0.5,0.5) {$T$};
\node at (-0.5,-1.5) {$R_1$};
\node at (-2.5,1.5) {$R_2$};
\node at (2.5,2.5) {$R_3$};
\end{tikzpicture}
}
}
\caption{\label{fig:partition} Different regions $R_j$ surrounding the target region $T$. We describe the macroscopic state of the surrounding of $T$ by specifying the fraction of occurrences of each symbol in $\Sigma$ for each $R_j$.}
\end{figure}

For some finite target region $T$ and a desired 
transformation $\beta: \Sigma^T \to \Sigma^T$ 
we want to initialize $\Lambda \setminus T$ in such a way that $\beta$ is implemented after some time $t\in \N$, but we demand that
only the `macroscopic' properties of some initial configuration $c \in \Sigma^\Lambda$ should matter. 
Due to the locality of update rules in CAs, only a finite region $R$ surrounding $T$ is relevant 
for the implementation anyway. 
We therefore need to describe the `macroscopic' properties only for the restriction $c_R$ of $c$ to $R$.
To this end, we partition $R$ into the disjoint union of
$m$ regions $R=R_1\cup R_2 \cup \cdots \cup R_m$
as shown in Figure~\ref{fig:partition}. One may think of  these regions as representing pieces of matter made of different material. 
\subsection{Classical CA}
Motivated by macroscopic observables like `mean magnetization', let us  specify the macroscopic state of any region $R_j$ as follows. For each $s_1,\dots,s_k\in \Sigma$, let $N^j_1(c_R),\dots,N^j_k(c_R)$ denote the numbers of occurences of symbols 
in $R_j$. Then define the corresponding densities via the normalization
$n^i_i(c_R):=N^j_i(c_R)/|R_j|$.
As sufficient condition for $c_R$ to implement $\beta$, we assume 
\begin{equation}\label{eq:constr}
|n_j^i(c_R) - l^i_j|\leq \epsilon,
\end{equation}
where $\epsilon$ is some error tolerance.
If we think of the regions $R_j$ to consist of a macroscopic number of cells (order of $10^{23}$ like Avogadro's constant, for instance), $\epsilon$ is thought to be much closer to $1$ than to
the inverse of this huge number -- otherwise we would not call the deviation `macroscopic'.
 
Let us now consider a configuration $c'_R$ that satisfies our macroscopic constraints
\eqref{eq:constr} better than required, that is, with $\epsilon/2$ instead of $\epsilon$.
We then consider the configuration $c''_R=\tau(c'_R)$, where $\tau$ denotes the right shift by one cell. Whenever the regions $R_j$ are large enough
to ensure that the overlap
$\tau(R_j) \cap R_j$ covers more than the fraction $1-\epsilon/2$ of $R_j$, the densities cannot change 
by more than $\epsilon/2$.
Thus,
$c''_R$ still satisfies \eqref{eq:constr}, which
guarantees that it also implements $\beta$.
Assume now that $T$ consists of the two adjacent cells
$(0,0,\cdots,0)$ and $(1,0,\dots,0)$. Further assume that it acts a NOT on the first cell and the identity on the second one.
However, if $c'_R$ implements $\beta$, translation invariance of the update rules imply that 
$c''_R$ implements $\tau \circ \beta \circ \tau^{-1}$.
Hence, $c''_R$ implements a NOT on
 $(1,0,\dots,0)$  although we have argued that it implements ID on that cell, as $c'_R$ does. 
We phrase this simple observation as a theorem:

\begin{Theorem}[no classical physically universal CA with macroscopic interface]
Given a CA with dimension $d\in \N$. 
Assume we describe the macroscopic state of
the complement of $T$ by densities of symbols in
some regions $R_j$ for which
\[
r_j:=\frac{|\tau(R_j) \cap R_j| - |R_j|}{|R_j|} \leq \epsilon/2. 
\]
Then there cannot be constants $l^i_j$ such that
every initialization
 satisfying \eqref{eq:constr} implements
the operation (NOT,ID) on the cells $(0,0,\cdots,0)$ and $(1,0,\cdots,0)$. 
\end{Theorem}

Although our specific notion of `macroscopic' may not necessarily be the right one for our purpose, our conclusion 
seems to be quite robust regarding reasonable redefinitions. 
Specifying the initial state 
in a way that is insensitive to shifting  it
by a few cells (which should be true for any macroscopic description), cannot implement any precisely localized operation.
In the terminology of  \cite{BRS06a,Referenz}, it is impossible to generate {\it reference information}  from zero, a notion that has meanwhile been defined within a quite general framework of resource theories.\footnote{\cite{clock}, for instance, introduces
a {\it quasi-order} of clocks in which
no time covariant operation can generate a clock that is more precise than the `resource' clock.}

Since the above shift operation can be seen as a translation of hardware of the entire control device, it is obvious that it simply shifts the target operation. Rather than demanding 
a target operation whose action is specified by an  {\it absolute} position in space, it thus seems more appropriate to define a target region relative to the position of the control device (in analogy to a relational formulation
of quantum theory
\cite{PoulinRel} in which only relative coordinates are accessible). The way to get out of this problem, however, is not as obvious as it seems. After all, no description of 
the hardware of a realistic control device would be precise enough to admit an obvious definition of its `location in space' up to the scale of single cells (which would be required for defining the relative position of the target region). To see this, think of variations of a hardware where the {\it shapes} of the regions $R_j$ differ on the scale of single cells. 
In some cases, it could be that only the position of the barycenter of
one specific $R_j$ matters for the position of the target operation. In others, it could be the barycenter over all regions, or also more sophisticated functions of the shape. To further elaborate on the idea that there is no obvious definition of relative position, think of the case where some regions $R_j$ are shifted by one cell, others remain fixed, and the remaining regions slightly change their shape to yield a valid partition.
What is the `position' of the new
device relative to the orginal one? Does it differ by one cell or not?

\subsection{Quantum CA}
We will see that there is nothing substantially different for an analog
argument for the quantum CA, although some steps need to be modified. For an analog definition
of `densities' on the regions $R_j$ we use the mean-field observables of the type \eqref{eq:meanfield}: 
 For any observable $a\in M(\C^{|\Sigma|})$ with
$\|a\|\leq 1$  we denote its copy on site $\lambda\in \Lambda$ by $a_\lambda$ and then define the mean field observable for any region $R$ by
\[
\overline{a}_{R}:= \frac{1}{|R|} \sum_{\lambda \in R} a_\lambda.
\]
While macroscopic observables in mathematical physics are usually defined via limits of averages
of infinitely growing regions \cite{BR2}, 
we want to explicitly stay ways from these `tricks'. This would be an idealization that could not provide convincing arguments for fundamental limits. 

For a state $\rho$ to implement our target operation $\beta$ we assume that all these mean field observables should attain values that are close to $l^{a}_j$ up to a small deviation only. We therefore require
\begin{equation}\label{eq:quantumcontraints}
\rho\left[(\overline{a}_{R_j} - l^a_j \one )^2\right] \leq \epsilon,
\end{equation} 
with $\one$ denoting the identity in $\cA_\Lambda$ and 
$l^a_j$ some constants in $[-1,1]$.
Note that different mean field observables on the same region $R$ commute up to an error term of operator norm $O(1/|R|)$, which is easily seen by straightforward counting of terms. For large regions, they are thus {\it almost} compatible. 
For product states, for instance, all mean field observables have uncertainty of the order $O(1/\sqrt{|R|})$. Therefore, it is for sufficiently large $R$, feasible to specify the values of all mean-field observables by constraints like \eqref{eq:quantumcontraints} with appropriate values $l^a_j$. 

Introducing the shift $\tau^*$ on $\cA_{\Lambda}$ simple counting arguments show 
\[
\| \tau^* (\overline{a}_{R_j}) - \overline{a}_{R_j}\| \leq 
r_j,
\]
with $r_j$ as in Theorem~\ref{thm:class}.
Hence, 
\[
\|\tau^* \left((\overline{a}_{R_j} - l^a_j \one)^2\right) -
 (\overline{a}_{R_j} - l^a_j \one)^2\| \leq 2 r_j. 
\]
Whenever we ensure that $\rho$ satisfies the bound
\eqref{eq:quantumcontraints} with $\epsilon/2$ instead of
$\epsilon$ and we ensure that the regions $R_j$ are large enough to satisfy $2r_j\leq \epsilon/2$, we thus know that
$\tau(\rho)$ still implements the same transformation as $\rho$.
Finally, we  obtain the following result -- which comes without any surprises:
\begin{Theorem}[no physically universal quantum CA with macroscopic interface]\label{thm:class}
Whenever the regions $R_j$ satisfy $r_j\leq \epsilon/4$,
it is not possible that all states $\rho$ whose mean field observables satisfy
the constraints \eqref{eq:quantumcontraints} implement
(NOT,ID) on the cells $(0,0,\dots,0)$ and
$(1,0,\cdots,0)$.
\end{Theorem}

\section{Conclusions}

Within our highly idealized model class, we have shown that microscopic actions cannot be achieved by controlling only the macroscopic state of the controller.
 We conclude that physically universal CAs only admit universal control of
microscopic degrees of freedom if one is able to act on the {\it microscopic} degrees of freedom of the   controller. In the strict sense of
physical universality defined previously,
macroscopic control of the controller is thus insufficient to achieve universality.
We have argued that it is necessary to rephrase universality in a sense that accounts for
the absence of an absolute spatial reference systems, but it is not obvious how to achieve this. 
The right class of models that describes how humans are able to act on microscopic systems by changing the state of macroscopic control devices by hand thus remains to be found.

Understanding this may help deciding whether any unitary can be implemented or whether there are fundamental limitations for 
many-particle systems. Even if such a theory
of the interface predicted the limitations that all elementary unitaries can only be implemented with some tiny inaccuracies, this could have dramatic implications for the appearance of a classical world. \cite{JB00}, for instance,
showed that macroscopic superpositions can no longer attained\footnote{This result is not challenged by the existence of error correcting codes since we care about physical states and not logical states that are embedded into a larger system.}. The idea that
a classical world emerges from decoherence is certainly standard. This insight would obtain, however, an even more fundamental significance
if one could show a universal 
microscopic theory of the
controlling interface itself entails 
limits of controllability.

\paragraph{Acknowledgements}
Thanks to Rob Spekkens for helpful discussions on the relation to reference frames.

\bibliographystyle{unsrt}
\bibliography{../../literatur/literatur}

\end{document}